
\magnification=1200
\overfullrule = 0 pt
\baselineskip=24 true bp
\hfuzz=0.5pt

\def\ie{{\it i.e.\/}}
\def\av#1{{\langle{#1}\rangle}}
\def\NI{\noindent}
\def\Dx{{\Delta x}}
\def\D2x{{(\Dx)^2}}
\def\tp{{{\tilde p}}}
\def\tG{{{\tilde G}}}

\def\erfc{{\rm erfc}}
\def\Ai{{\rm Ai}}
\def\cor{{c^{(2)}}}
\def\tcor{{{\tilde c}^{(2)}}}

\centerline{\bf Inter-Particle Distribution Functions for One-Species}
\centerline{\bf Diffusion-Limited Annihilation, $A+A\to 0$}
\vskip 0.8cm
\centerline{Pablo A. Alemany\footnote{$^{\dag}$}
{e-mail: {\sl alemany@cab.cnea.edu.ar} and
{\sl alemany@tpoly.physik.uni-freiburg.de}} }
\medskip
\centerline{Instituto Balseiro, 8400--Centro At\'omico Bariloche,
Argentina}
\centerline{and}
\centerline{Theoretische Polymerphysik, Universit\"at Freiburg,}
\centerline{Rheinstr. 12, D-79104 Freiburg, Germany}
\bigskip
\centerline{Daniel ben-Avraham\footnote{$^{\ddag}$}
{e-mail: {\sl qd00@craft.camp.clarkson.edu}} }
\medskip
\centerline{Physics Department, Clarkson University,
Potsdam NY 13699--5820, USA}
\centerline{and}
\centerline{Center for Nonlinear Studies, MS-B258}
\centerline{Los Alamos National Laboratory}
\centerline{Los Alamos, NM 87545}
\vskip 0.8cm
\NI{\bf ABSTRACT:}
Diffusion-limited annihilation, $A+A\to 0$, and coalescence, $A+A\to A$, may
both be exactly analyzed in one dimension.  While the concentrations of $A$
particles in the two processes bear a simple relation, the inter-particle
distribution functions (IPDF) exhibit remarkable differences.  However, the
IPDF is
known exactly only for the coalescence process. We obtain the IPDF for the
annihilation process, based on the Glauber spin approach and assuming that the
IPDF's of nearest-particle pairs are statistically independent. This
assumption is supported by computer simulations.  Our analysis
sheds further light on the relationship between the annihilation and the
coalescence models.
\medskip
\NI PACS:\ \ 82.20.Mj,\ \ 05.70.Ln,\ \  02.50.--r,\ \ 68.10.Jy,
\vfill\eject

\NI{\bf 1. Introduction}
\medskip\NI
One-species annihilation, $A+A\to 0$, and coalescence,
$A+A\to A$, have been extensively studied as basic prototypes of
diffusion-limited reactions [1-21].  In one dimension, these simple models
yield themselves to exact analysis: exact results for
coalescence have been obtained, for example,
with the method of inter-particle distribution
functions (IPDF) [12-18], and annihilation has been analyzed exactly
through a Glauber spin formalism [19-21] (and by several other techniques).

The distribution of distances between nearest particles, \ie, the IPDF,
plays a fundamental role in the Smoluchowski theory of diffusion-limited
reactions
and it has been studied extensively for various
reaction models [21-30].  In the case of one-species coalescence in one
dimension
the IPDF is obtained exactly, as a byproduct of the IPDF method
[15]. For one-species annihilation in one dimension, several aspects of
the kinetics are
known exactly, including the time dependence of the concentration of particles,
but not the IPDF.

In this paper, we compute the IPDF for the annihilation process from the
(exactly known) spin-spin correlation function of the Glauber spin formalism.
Our derivation is based on the assumption that the IPDF's of different particle
pairs are statistically independent.
Although this may seem a strong assumption,
comparison to computer simulations data suggests
that it may be exact.
At the very least, our approach yields a remarkably good
analytic approximation for the IPDF of the one-species annihilation model.

The rest of this paper is organized as follows.  In the next section
we briefly summarize the exact results known for the annihilation and the
coalescence processes.  We also discuss the similarities between
the IPDF method and the Glauber spin formalism.  In Section~3, we develop
a relation between the spin-spin correlation function in the Glauber method and
the IPDF of the annihilation process.  In Section~4, we apply this technique
to the computation of the IPDF for
annihilation, and for annihilation with input, and compare to results from
numerical simulations.  In the discussion, in Section~5, we reexamine
the conjecture of statistical independence.  We demonstrate the need for
proof by showing that the conjecture is wrong in the case of the closely
related
model of coalescence.

\bigskip
\NI{\bf 2. Coalescence and Annihilation}
\medskip\NI
Our models are defined on a one-dimensional lattice with lattice spacing
$\Dx$.
Each lattice site can be either empty, or occupied with exactly one particle.
Particles hop to the nearest site to their right or left at rate
$D/\D2x$.
On long length and time scales this yields normal diffusion with diffusion
coefficient $D$.  When a particle hops onto a site which is already occupied,
a reaction takes place: in the case of coalescence the impinging
particle is removed, modeling the reaction $A+A\to A$, and in the case of
annihilation both particles (the hopping particle and the target) are removed,
modeling $A+A\to 0$.

Coalescence has been analyzed exactly by the IPDF method (also known as the
method of empty intervals) [see ref.~15 for a review].  In this approach,
one defines $E_n(t)$ as the probability that an arbitrary sequence of $n$
consecutive sites is empty at time $t$.  From the $E_n$ one can compute various
quantities of interest.  For example, the concentration of particles is given
by
$$
c(t)={1-E_1(t)\over\Dx}. \eqno(1)
$$
The $E_n$ satisfy a simple, closed master equation which can be solved exactly
not only for the basic coalescence model, but also for a number of
variations, including the back reaction, $A\to A+A$, input of $A$ particles,
inhomogeneous initial distributions, and finite-size lattices [11-18].

Although the analysis can be carried out in discrete form, it is much
simpler (and more transparent) to consider the continuum limit.  This is done
by defining $x=n\Dx$ and then letting $\Dx\to 0$.  The empty interval
probabilities are replaced by the two-variable functions, $E(x,t)$, and the
concentration, for example, becomes
$$
c(t)=-{\partial E(x,t)\over\partial x}|_{x=0}. \eqno(2)
$$
The IPDF $p(x,t)$,
\ie, the probability that the nearest particle to an arbitrary particle
is a distance $x$ away at time $t$, can also be obtained from $E(x,t)$:
$$
c(t)p(x,t)={\partial^2 E(x,t)\over\partial x^2}. \eqno(3)
$$
Throughout the remainder of this paper we will confine ourselves to the
continuum limit.

In the case of simple coalescence
the equation of motion for the empty interval probability is
$$
{\partial E\over\partial t}=2D{\partial^2 E\over\partial x^2}, \eqno(4)
$$
with boundary conditions
$$
E(0,t)=1\qquad {\rm and}\qquad E(\infty,t)=0. \eqno(5)
$$
For generic homogeneous initial conditions (excluding exotic
fractal distributions of the particles [11]) the system arrives at a universal
long time asymptotic regime, where
$$
c(t)\to {1\over\sqrt{2\pi Dt}},\qquad {\rm as\ \ }t\to\infty, \eqno(6)
$$
and
$$
p(x,t)\to {x\over 4Dt}\exp(-{x^2\over 8Dt}),
\qquad {\rm as\ \ }t\to\infty.  \eqno(7)
$$
The IPDF can be put in scaling form in terms of
the dimensionless interparticle distance $\xi=c(t)x$:
$$
p(\xi,t)=c(t)p(x,t)\to {\pi\over 2}\xi\exp(-{\pi\over 2}{\xi^2\over 2}),
\qquad {\rm as\ \ }t\to\infty.  \eqno(8)
$$
Notice that in the long time asymptotic limit it becomes stationary.

The annihilation model is most easily analyzed in terms of its dual Glauber
spin process (in the zero temperature limit) [19,21].
In the latter, each lattice site is occupied by a `spin'
variable which assumes one of two states, $\sigma=\pm 1$.  Any spin adjacent to
a spin of opposite state may change its state, at rate $D/\D2x$.
If we associate
a particle with each pair of anti-parallel spins (\ie, the particles are
at the boundaries between alternating domains of parallel spins, Fig.~1), these
particles are seen
to diffuse and annihilate upon encounter, exactly as in the annihilation model.

Consider the spin-spin correlation function,
$G(x',x'',t)=\av{\sigma(x',t)\sigma(x'',t)}$, where
the angular brackets denote an average over different realizations of the
process. For our purpose it is sufficient to limit the discussion to
homogeneous
initial distributions of `particles', in which case $G(x',x'',t)$ is
a function of $x=x''-x'$.
The correlation function, $G(x,t)$, satisfies the diffusion equation [19,21]
$$
{\partial G\over\partial t}=2D{\partial^2G\over\partial x^2}, \eqno(9)
$$
with boundary conditions
$$
G(0,t)=1\qquad {\rm and}\qquad G(\infty,t)=0. \eqno(10)
$$
The concentration of particles is obtained from the correlation function, as
$$
c(t)=-{1\over 2}{\partial G(x,t)\over\partial x}|_{x=0}. \eqno(11)
$$
Notice that $G$ satisfies the same equation as $E$, and for generic initial
conditions it attains the same asymptotic limit as $E$.  It follows that
the concentration of particles for annihilation is exactly one half of
the concentration in the coalescence process (in the long time asymptotic
limit).  In fact, the two processes maintain a similar simple relation
at all times [3].

Another approach to the annihilation model relies on $P_e(x,t)$, the
probability that there is an even number of particles in an arbitrary interval
of length $x$ at time $t$ [20].
$P_e$ bears a simple relation to the spin-spin correlation
function: $G=2P_e-1$, and hence it too satisfies Eq.~(9), but with
$P_e(\infty,t)=1/2$.  The advantage of this approach is in that it deals
with the particles directly and it does
not require the dual process with spins, but it is essentially the same.

In spite of these exact analyses, there is no obvious way to determine the
IPDF for the annihilation process.  Computer simulations reveal that the long
distance tail of the IPDF falls off exponentially, as $e^{-x}$ [15,21].  This
is also supported by non-rigorous theoretical arguments [15].  In this
respect the annihilation process is radically different from coalescence
(compare Eq.~8).  In the following, we develop a method for calculating the
IPDF in the annihilation process.

\bigskip
\NI{\bf 3. Relation between IPDF and Correlation Functions}
\medskip\NI
For the annihilation process, the IPDF may be related to the spin-spin
correlation function by assuming that at any particular time
the particles (or the spin flips)
constitute a renewal process on the line [31].  The IPDF is the the renewal
probability density.
For simplicity, we shall omit the time variable, since it plays no role in
the spatial renewal process of the particles.

Let $p_{\parallel}(x)$ be the probability that two spins separated by a
distance $x$ are parallel.  Then, the spin-spin correlation function is
simply
$$
G(x)=p_{\parallel}-(1-p_{\parallel})=2p_{\parallel}(x)-1. \eqno(12)
$$
Two spins would be parallel if there are no spin flips in between them,
or if there is an even number of spin flips in between them.  (Indeed,
$p_{\parallel}$ is identical to $P_e$.)
Let $p_0(x)$
be the probability that the first flip from a given spin occurs at a distance
$x$ from it (the forward waiting distribution function). $p_0$ is
related to the renewal probability distribution (the IPDF, $p$) through
[32]:
$$
p_0(x)=\int_x^{\infty}{{cx'p(x')\,dx'\over x'}}=
c\int_x^{\infty}{p(x')\,dx'},  \eqno(13)
$$
where $c=1/\int_0^{\infty}xp(x)\,dx$ is a normalization constant which
happens
to be equal to the concentration of particles.
The probability that there are no spin flips up to a distance $x$ from a given
spin is
$$
p_0'(x)=\int_x^{\infty}{p_0(x')\,dx'}.  \eqno(14)
$$
(Notice that $p_0'$ is identical to the empty interval probability $E$, defined
for the coalescence process.)
Finally, let $p_1(x)$ be the probability
that there are no spin flips up to a distance $x$ from a given spin flip:
$$
p_1(x)=\int_x^{\infty}p(x')\,dx'={p_0(x)\over c}. \eqno(15)
$$

We can now express $p_{\parallel}$ as the sum of $p_0'$, and $p_0$ convoluted
with $p$ an {\it odd\/} number of times (this yields an even number of spin
flips) and convoluted with $p_1$.  Working with the Laplace transform of the
distribution functions (which we shall denote by a tilde), the convolutions
assume the simpler form of products:
$$
\tp_{\parallel}= \tp_0'+\tp_0(\tp+\tp^3+\cdots)\tp_1=
{1\over s}-{c\over s^2}{1-\tp\over 1+\tp}. \eqno(16)
$$
Here $s$ is the Laplace transform variable (conjugate to $x$), and the
last equality has been obtained using Eqs.~(13)-(15) and standard properties
of the Laplace transform.  From
Eq.~(12), we finally obtain the relation between the spin-spin correlation
function and the IPDF:
$$
\tp(s)={1-h(s)\over 1+h(s)};\qquad
h(s)\equiv {s\over 2c}\big(1-s\tG(s)\big).  \eqno(17)
$$

Assuming statistical independence of consecutive IPDF's one can also derive
an expression for the density-density correlation function $\cor(x,t)$, \ie,
the joint probability of finding any two particles separated by a distance $x$
at time $t$.
Because there can be any integer number of particles between two arbitrary
particles,
the density-density correlation function is given by  a sum over
convolutions of the IPDF.
In terms of the Laplace transform:
$$
{1\over c}\tcor=\tp+\tp^2+\tp^3+\cdots={\tp\over\ 1-\tp}={1-h(s)\over 2h(s)}.
\eqno(18)
$$

\bigskip
\NI{\bf 4. IPDF's for the Annihilation Process}
\medskip\NI
Let us now apply the above formalism to annihilation in the long time
asymptotic limit. In this case, for generic initial conditions the correlation
function approaches a universal form:
$$
G(x,t) = \erfc({x\over 2\sqrt{2Dt}}) \qquad{\rm as\ \ }t\to\infty, \eqno(19)
$$
and the concentration reaches the asymptotic limit
$$
c(t)={1\over 2\sqrt{2\pi Dt}}, \qquad{\rm as\ \ }t\to\infty. \eqno(20)
$$
With these expressions, we get from Eq.~(17),
$$
\tp={1-\sqrt{\pi}ks\,e^{k^2s^2}\erfc(ks)
\over 1+\sqrt{\pi}ks\,e^{k^2s^2}\erfc(ks)},  \eqno(21)
$$
where $k=\sqrt{2Dt}$.  This can be inverted either numerically or through
series expansions.  For small $x$ we find (in scaling form, with $\xi=cx$)
$$
{1\over c}p=\pi\xi-{5\over 6}\pi^2\xi^3+{49\over 120}\pi^3\xi^5+\cdots
\qquad (\xi\ll 1). \eqno(22)
$$
The linear term has been derived exactly by Amar and Family [21].
For large $x$ we get the asymptotic form (by contour integration)
$$
{1\over c}p\sim {4\sqrt{\pi}a\over 4a^2+1}e^{-2\sqrt{\pi}a\xi},
\qquad (\xi\gg 1),  \eqno(23)
$$
where $a=-z_0\approx 0.357835$ is the absolute value of the smallest negative
root of $1+\sqrt{\pi}z\exp(z^2)\erfc(z)=0$.  Thus, we do confirm the large
distance exponential decay of the IPDF.

We have performed a similar analysis for the transient
regime at early times.  Here, a discrete formulation is more appropriate, but
that is derived in a straightforward, analogous way to the continuum limit
of Section~3 [31]. Some typical results are shown in Figs.~2(a) and (b).
In Fig.~2(c) we compare simulation results at a very late time, well after the
long time asymptotic limit sets in, to the analytical IPDF obtained
from Eq.~(21).  There is perfect agreement between theory and simulations,
to the level of the statistical noise.  Notice, however, that
it is difficult to obtain good statistics in the long time asymptotic limit,
due to the small number of surviving particles.

We next analyze annihilation with a steady homogeneous input of particles
at rate $R$ per unit time per unit length.  This model has the advantage of
a stationary state, making it possible to obtain exceptionally good
statistics for the IPDF from simulations.  In this case, the spin-spin
correlation function satisfies the equation [19]
$$
{\partial G\over\partial t}=2D{\partial^2G\over\partial x^2}-2RG, \eqno(24)
$$
with the stationary solution
$$
G(x)={\Ai(r^{1/3}x)\over\Ai(0)}, \eqno(25)
$$
($r\equiv R/D$).  The stationary concentration is
$$
c={|\Ai'(0)|\over 2\Ai(0)}r^{1/3}, \eqno(26)
$$
where the prime denotes differentiation of the Airy function.

In Fig.~3 we compare simulation results to the analytic IPDF obtained from
Eqs.~(17), (25) and~(26).  Again, the agreement is excellent.  The slight
discrepancy at the peak of the curve may be a result of discreteness. Indeed,
this discrepancy becomes smaller for smaller steady state concentrations.
For short distances, we get $(1/c)p(\xi)=4(\Ai(0)/|\Ai'(0)|)^3\xi$.
For long distances,
we find an exponential decay of the
IPDF tail, similar to pure annihilation.
This is interesting, in view of the fact that the spin-spin correlation
function decays as $\exp(-x^2)$ for pure annihilation, but as $\exp(-x^{3/2})$
in the case of input.

\bigskip
\NI{\bf 5. Discussion}
\medskip\NI
Our analytical approach would be exact if the IPDF's for
different particle pairs in the annihilation process were statistically
independent.  Only then could the convolutions leading to Eqs.~(17) and
(18) be justified.  Therefore, the excellent agreement between simulations
and the analytical computations leads us to the following conjecture:
\smallskip
\NI{\sl
The distribution functions of the inter-particle distances of different
particle pairs in
the annihilation process are statistically independent from each other.}
\smallskip
We assume, of course, that at time $t=0$ the initial distribution of particles
is such that IPDF's are statistically independent.  If such is the case,
our results suggest that the
statistical independence would be maintained at all times.

It is not obvious why this conjecture should be true.  In fact, we now
show that in the coalescence process the IPDF's of different particle
pairs are statistically {\it dependent\/}.
The density-density correlation function for the coalescence process
has been obtained exactly [33, 34]:
$$
\cor(\xi)=c^2[1-e^{-{\pi\over 2}\xi^2}
+{\pi\over 2}e^{-{\pi\over 4}\xi^2}\erfc({\sqrt{\pi}\over 2}\xi)]. \eqno(27)
$$
On the other hand, we can compute $\cor$ assuming statistical independence,
from Eqs.~(8) and (18).  In spite of a rough qualitative agreement, there are
important discrepancies: for example,
the exact result of Eq.~(27) shows that
$\cor$ approaches the long distance limit as
$c^2-\cor\sim\exp(-{\pi\over 2}\xi^2)$,
monotonously, while statistical independence predicts an exponential
approach with oscillations, $c^2-\cor\sim\exp(-a\xi)\cos(b\xi+\phi)$,
($a$, $b$, and $\phi$ are known constants [34]).

This poses the following puzzle.  Consider diffusion-limited polymerization of
$n$-mers, $A_n$.
The polymers diffuse on the
line, with a diffusion constant independent of their size,
and polymerize upon encounter: $A_i+A_j\to A_{i+j}$. Initially, all particles
are monomers, $A_1$.  This process
codes for coalescence and for annihilation simultaneously [10]:
If we disregard size, and
consider all polymers as equivalent particles, the process is analogous
to coalescence, $A+A\to A$.
On the other hand, if we focus only on polymers of an {\it odd\/} number of
monomers, the
process is equivalent to annihilation, $A+A\to 0$.  Imagine now that we
start this ``master-process" and we look at a snapshot of the system at some
later time.  There are polymers of odd and even size, roughly in
equal amounts, and apparently well mixed (there is no segregation into
clusters of odd and even sized polymers).  From our foregoing discussion we
conclude that the distances between any pair of nearest polymers
are statistically dependent (we are looking at coalescence).  But if our
conjecture is true, the distances between nearest polymers of odd
size (annihilation)---a finite subset of all polymers---are statistically
independent!
There is of course no contradiction, however, the spatial distribution of the
polymers in the master-process must then be rather peculiar.

In conclusion, we have presented strong numerical evidence that the
distribution functions of distances between nearest particle pairs
in the annihilation
process, and in the annihilation process with input, are statistically
independent.  If we accept this as true, then the IPDF for the annihilation
model has been computed exactly.  Our computations reproduce the exact short
distance limit of Amar and Family [21].  We also confirm the exponential
decay of the IPDF tail.  The IPDF for the case of annihilation with input
is similar to that of pure annihilation, but the depletion zone near the
origin is narrower.  This is the result of the input, which tends to bring
the IPDF closer to a pure exponential, because of its random nature.

Our work suggests a most significant difference between annihilation
and coalescence:  the IPDF's are statistically independent for the former,
but not for the latter.  An interesting result is that the density-density
correlation function exhibits oscillations in the case of annihilation,
but not for coalescence.  Such oscillations are typical of fluids with
hard core repulsion interactions, and are expected because of the effective
repulsion between particles due to the reactions.  Their absence in the
case of coalescence is a striking consequence of the correlations between
IPDF's.

The excellent agreement between our analytical derivations and simulations
suggests that our approach is exact.  We have not yet been able to
prove the necessary conjecture of statistical independence (for the
annihilation
model).  This remains
an open problem.

\bigskip
\NI{\bf Acknowledgments}
\medskip\NI
This work was partially funded by a grant from Fundaci\'on Antorchas
(Argentina) and the Deutsche Forschungsgemeinschaft (Grant No.~SFB~60).
We are grateful to the Theoretische Polymerphysik at Freiburg
University and the Physics Department at Clarkson University (PAA),
and to the Center for Nonlinear Studies at the Los Alamos National
Laboratory (DbA) for their hospitality while this work was in progress.
We thank Charlie Doering for interesting, useful discussions, and we thank
M. Burschka, C. Doering, and W. Horsthemke for sharing their results with
us prior to publication.

\vfill\eject
\centerline{\bf References}
\medskip

{\frenchspacing

\item{1.} K.~Kang and S.~Redner,
{\sl Phys. Rev.} A {\bf 32}, 435 (1985).

\item{2.} V.~Kuzovkov and E.~Kotomin,
{\sl Rep. Prog. Phys.} {\bf 51}, 1479 (1988).

\item{3.} V. Privman, {\sl J. Stat. Phys.} {\bf 69}, 629
(1992): {\sl J. Stat. Phys.} {\bf 72}, 845 (1993).

\item{4.} G. Zumofen, A. Blumen, and J. Klafter, {\sl J. Chem. Phys.}
{\bf 82}, 3198 (1985).

\item{5.} R. Kopelman, {\sl J. Stat. Phys.} {\bf 42}, 185 (1986);
{\sl Science} {\bf 241}, 1620 (1988).

\item{6.} M. Bramson and D. Griffeath, {\sl Ann. Prob.} {\bf 8}, 183 (1980);
{\sl Z. Wahrsch. Geb.} {\bf 53}, 183 (1980).

\item{7.} D. C. Torney and H. M. McConnell, {\sl J. Phys. Chem.} {\bf 87},
1941 (1983).

\item{8.} L. Peliti, {\sl J. Phys. A} {\bf 19}, L365 (1985).

\item{9.} A. A. Lushnikov, {\sl Phys. Lett. A} {\bf 120}, 135 (1987).

\item{10.} J. L. Spouge, {\sl Phys. Rev. Lett.} {\bf 60}, 871 (1988).

\item{11.} P. A. Alemany and D. H. Zanette, {\sl Chaos, Solitons and
Fractals} {\bf 6}, 7 (1995).

\item{12.} C. R. Doering and D. ben-Avraham, {\sl Phys. Rev. A} {\bf 38},
3035 (1988).

\item{13.} C. R. Doering and D. ben-Avraham, {\sl Phys. Rev. Lett.} {\bf 62},
2563 (1989).

\item{14.} M. A. Burschka, C. R. Doering, and D. ben-Avraham, {\sl Phys.
Rev. Lett.} {\bf 63}, 700 (1989).

\item{15.} D. ben-Avraham, M. A. Burschka, and C. R. Doering,
{\sl J. Stat. Phys.} {\bf 60}, 695 (1990).

\item{16.} C. R. Doering and M. A. Burschka, {\sl Phys. Rev. Lett.} {\bf 64},
245 (1990).

\item{17.} C. R. Doering, M. A. Burschka, and W. Horsthemke, {\sl J. Stat.
Phys.} {\bf 65}, 953 (1991).

\item{18.} C. R. Doering, {\sl Physica A} {\bf 188}, 386 (1992).

\item{19.} Z. R{\'a}cz, {\sl Phys. Rev. Lett.} {\bf 55}, 1707 (1985).

\item{20.} H. Takayasu, I. Nishikawa, and H. Tasaki, {\sl Phys. Rev. A}
{\bf 37}, 3110 (1988).

\item{21.} J. G. Amar and F. Family, {\sl Phys. Rev. A}, {\bf 41},
3258 (1990); M. M\"uller and W. Paul, {\sl Europhys. Lett.}
{\bf 25}, 79 (1994).

\item{22.} G. H. Weiss, R. Kopelman, and S. Havlin, {\sl Phys. Rev. A}
{\bf 39}, 1620 (1989).

\item{23.} D. ben-Avraham and G. H. Weiss, {\sl Phys. Rev. A} {\bf 39},
466 (1989).

\item{24.} H. Taitelbaum, R. Kopelman, G. H. Weiss, and S. Havlin,
{\sl Phys. Rev. A} {\bf 41}, 3116 (1990).

\item{25.} S. Havlin, H. Larralde, R. Kopelman, and G. H. Weiss,
{\sl Physica A} {\bf 169}, 337 (1990).

\item{26.} S. Redner and D. ben-Avraham, {\sl J. Phys. A} {\bf 23},
L1169 (1990).

\item{27.} R. Schoonover, D. ben-Avraham, S. Havlin, R. Kopelman,
and G. H. Weiss, {\sl Physica A} {\bf 171}, 232 (1991).

\item{28.} S. Havlin, R. Kopelman, R. Schoonover, and G. H. Weiss,
{\sl Phys. Rev. A} {\bf 43}, 5228 (1991).

\item{29.} G. H. Weiss and J. Masoliver, {\sl Physica A} {\bf 174},
209 (1991).

\item{30.} G. H. Weiss, {\sl Physica A} {\bf 192}, 617 (1993).

\item{31.} P. A. Alemany: {\sl Ph. D. Dissertation}, Instituto
Balseiro, Centro At\'omico Bariloche, Argentina, and Theoretische
Polymerphysik, Universit\"at Freiburg, Germany (1995).

\item{32.} see for example: D. R. Cox, {\sl Renewal Processes}, in Monographs
on
Statistics and Applied Probability, D. R. Cox and D. V. Hinkley eds.,
(Chapman and Hall, London, 1982).

\item{33.} M. A. Burschka, C. R. Doering, and W. Horsthemke, in preparation.

\item{34.} For a recent review on exact results about the coalescence process,
see: D. ben-Avraham, ``The Coalescence Process, $A+A\to A$, and the
Method of Interparticle Distribution Functions", in {\sl Nonequilibrium
Statistical Physics in One Dimension}, V.~Privman ed., (Cambridge
University Press, to appear).

}

\vfill\eject
\noindent
\centerline{\bf CAPTIONS}
\bigskip
\bigskip
\bigskip

\NI\hang{\bf Figure 1:} Relation between the Glauber spin model and the
annihilation process.  The $A$ particles (bottom) correspond to the spin
flips, or domain boundaries, in the spin process (top).
\bigskip

\NI\hang{\bf Figure 2:} Comparison of simulated IPDF's to
analytical derivations for the annihilation process, $A+A\to 0$.
Simulation results are shown in histogram form
for (a) $t$=1, (b) $t=10$, and (c) $t=1000$.  All simulations began with a
homogeneous density of particles, $1/2$.
\bigskip

\NI\hang{\bf Figure 3:} Comparison of the simulated IPDF (circles) to
the analytical derivation (solid curve) for the annihilation process
with input.  The input rate ($R=0.0001$) is low enough so that the continuum
limit applies ($c\approx 0.0213\ll 1$).  Simulation results represent an
average
over $16\times 10^6$ time units (performed after the steady state was reached).
\vfill\eject
\bye